\documentclass[conference]{IEEEtran}
\usepackage{cite}
\usepackage{amsmath,amssymb,amsfonts}
\usepackage{graphicx}
\usepackage{textcomp}
\usepackage{url}
\usepackage{etoolbox}
\usepackage{cuted}
\usepackage{lipsum}
\usepackage{array}
\usepackage{color}
\usepackage{xcolor}
\usepackage{dirtytalk}
\usepackage{algpseudocode}
\usepackage{algorithm}
\usepackage[draft]{todonotes}
\usepackage[english]{babel}
\usepackage[utf8x]{inputenc}
\usepackage{amsmath,amsthm}

\usepackage{comment}
\usepackage{multirow}

\theoremstyle{definition}

\theoremstyle{plain}
\newtheorem{theorem}{Theorem}

\ifCLASSINFOpdf
\else
\fi
\begin{document}

%\title{Privacy-enhancing Verification for Dynamic Truck Platooning Based on Blockchain and Zero-knowledge Proof}

\title{Location-aware Verification for Autonomous Truck Platooning Based on Blockchain and Zero-knowledge Proof}

\author{\IEEEauthorblockN{Wanxin Li\IEEEauthorrefmark{1}~~~~Collin Meese\IEEEauthorrefmark{1}~~~~Zijia (Gary) Zhong\IEEEauthorrefmark{2}~~~~Hao Guo\IEEEauthorrefmark{3}~~~~Mark Nejad\IEEEauthorrefmark{1}}
\IEEEauthorblockA{\IEEEauthorrefmark{1}
Department of Civil and Environmental Engineering, University of Delaware, U.S.A.\\\IEEEauthorrefmark{2}John A. Reif, Jr. Department of Civil and Environmental Engineering, New Jersey Institute of Technology, U.S.A.\\\IEEEauthorrefmark{3}School of Software, Northwestern Polytechnical University, Taicang Campus, China.
\\
\{wanxinli,cmeese,nejad\}@udel.edu, zijia.zhong@njit.edu, haoguo@nwpu.edu.cn}

%\vspace{-.5 cm}
%{Authorship will be discussed and determined by contributions.}
}

\IEEEoverridecommandlockouts
\IEEEpubid{\makebox[\columnwidth]{978-0-7381-1420-0/21/\$31.00~\copyright2021 IEEE \hfill} \hspace{\columnsep}\makebox[\columnwidth]{ }}

\maketitle

%\IEEEpubidadjco

\begin{abstract}
Platooning technologies enable trucks to drive cooperatively and automatically, which bring benefits including less fuel consumption, more road capacity and safety. In order to establish trust during dynamic platoon formation, ensure vehicular data integrity, and guard platoons against potential attackers, it is pivotal to verify any given vehicle's identity information before granting it access to join a platoon. %In addition, truck owners may be reluctant to disclose private identifying information to a centralized entity or untrusted third parties due to privacy concerns. 
To address this concern in dynamic truck platooning, we present a novel location-aware and privacy-preserving verification protocol based on zero-knowledge proof and permissioned blockchain. By performing the verification process within the spatially-local area defined by a given platoon, our system can provide lower latency and communication overhead compared to a location-agnostic blockchain system. %Platoon records are stored directly on the blockchain ledger to guarantee the immutability and integrity of the data. Programmable access control policies enable truck companies to define who is allowed to access their platoon records. 
We prototype the proposed system and perform benchmark tests on the Hyperledger platform. The experimental results show that our system is suitable for real-world truck platooning. 

\end{abstract}


% \fi
\IEEEpeerreviewmaketitle

\section{Introduction}
%Truck platooning, enabled by  Vehicular Ad-Hoc Networks (VANET), is expected to affect the freight industry profoundly. % based on numerous studies. 
Truck platooning involves linking two or more trucks together in a convoy with a short following headway with wireless connectivity and vehicle automation. %By sharing control parameters (e.g., speed, direction, acceleration) amongst the vehicles, platoon members are able to achieve cooperative automated driving that results in less fuel consumption, greater roadway capacity, and, most importantly, safer operation. More specifically, a platoon member is able to safely follow its preceding vehicle at a much shorter headway (e.g., 0.6 seconds) compared to the conventional two-second following headway for a human driver. The drastic headway reduction could yield a lane capacity of 4,250 vehicles per hour\cite{van2006impact}, double the existing lane capacity. 
As a by-product of the short following headway, fuel efficiency is expected to improve. According to Japan ITS Energy project, 15\% of fuel can be saved with a  4.7-m intra-platoon following gap at 80 km/h \cite{tsugawa2014results}. Truck platooning also allows the driver to disengage from driving tasks. Human error was estimated to be responsible for 94\% of the traffic accident in the U.S. \cite{nhtsa2015Critical}. Compared to human drivers, automated driving systems could achieve a much shorter response time and more accurate assessment of the dynamic traffic conditions. %Lastly, truck platooning even allows the driver to undertake other non-driving tasks, such as administrative work.

% I moved merged the paragraph in the related work here
%There are two models for truck platooning: the road train model and the opportunistic model\cite{logistics2019}. The former involves platooning multiple trucks having the same origins and destinations. The latter aligns more with the VANET structure, where trucks from different fleets would need a mechanism to identify platooning opportunities and fairly appropriate platooning roles. To take full advantage of the estimated 45 thousand platoon-able miles in the U.S., an opportunistic model is necessary to accommodate heterogeneous fleets \cite{fleetowner2018} and serve the near \$600 billion US trucking market\cite{kearney2016accelerating}. However, it is still unclear how to integrate the opportunistic model under the context of a ``mixed fleet'' environment, where a platoon encompasses vehicles from different companies operating on distinctly heterogeneous and isolated systems. Mixed fleet platooning was demonstrated successfully in the 2011 Grand Cooperative Driving Challenge \cite{ploeg2012introduction}, but security and privacy were not the upmost concern due to the nature of the competition.

% It is believed that truck platooning will be first deployed by large fleet companies (e.g., FedEx, UPS), who benefit from economies of scale.

%, homogeneous fleets, and scheduling authority of their fleets. 

The security of the platooning systems, which protects them from unauthorized access to proprietary information about a specific vehicle or fleet specifications, under mixed fleet scenario has not been extensively studied thus far, though awareness of such aspect has gained increasing attention\cite{truckinginfo2020, chen2020smart, singh2020integrating}. To ensure the security of the system, it is crucial to be able to verify a given vehicle's identity information prior to granting it access to join a platoon in order to establish trust, ensure the platoon integrity and guard against potential attackers. %At the same time, CAV owners may be reluctant to disclose private identifying information to an untrusted third party due to privacy concerns. Moreover, if the dynamic platoon formation service is controlled by a single centralized entity, the system becomes vulnerable and can be at risk for data breaches and other attacks which can expose private user data. As a result, safeguarding the privacy of CAV owners while simultaneously providing a means of dynamic identifier verification for other heterogeneous CAVs, presents a pivotal challenge to realizing the benefits of dynamic truck platooning systems in a mixed fleet environment. 
 
%  Chen et al.\cite{chen2020smart} developed a smart-contract-based framework that aims to fairly apportion the fuel saving between platoon leaders and members.  Singh et al. \cite{Singh2020integrating} proposed a platoon management framework that is comprised of  a proof-of-work blockchain at the RSU level and a premissioned blockchain at vehicle level.

% Currently, we have seen driving features enabled by SAE Level 2 automation in the production vehicle in the form of standard or optional packages, including adaptive cruise control, traffic jam assist, pre-collision mitigation. Strictly speaking, the trucking platoon only pertains to the longitudinal control of a vehicle and therefore is classified as SAE Level 1 automation.  However, the communication layer of the truck platoon is more complex than other VANET-enabled applications. 

% There are various ways for platoon formation in dynamic traffic, including ad-hoc formation, local coordination, and global coordination (a.k.a. end-to-end platooning) \cite{zhong2020influence}. 
% Regardless of the criteria of platoon formation, 

Over the past decade, research in blockchain technology has highlighted it as a promising technology for supporting a myriad of decentralized applications between both trusted and anonymous peers \cite{wanxin2019blockchain, li2020blockchain, 8946149, 9169395, 9210529, 9217515}. In relation to dynamic truck platooning, blockchain presents some desirable properties for creating a robust, dynamic and decentralized system. %Succinctly, blockchain technology is a novel network design which enables consensus among distinctly heterogeneous network peers on a distributed and immutable ledger, and was first proposed in Bitcoin: A Peer-to-Peer Electronic Cash System \cite{nakamoto2008bitcoin}. %In general, blockchain technology comes in two forms: permissionless, where the network is open to anyone and users have pseudonymity, and permissioned, where network access must be granted and all participants are known \cite{liu2019will}. Both flavors of blockchain technology inherently provide properties of immutability, network reliability and data provenance, making them suitable candidates for supporting a dynamic truck platooning system. 

In this paper, we propose and prototype a system for identity verification in the context of dynamic truck platooning, motivated by permissioned blockchain technology and zero-knowledge proofs. %This paper makes the following scientific contributions:

\section{System Architecture}
%Components: Blockchain, CAV, Peer Node, Companies, Authority
%This section describes the proposed location-aware system architecture for identity verification in truck platooning, which includes the location-aware verification protocol, and the permissioned blockchain network with access control policies. 
By referring to Fig. \ref{fig:arch}, we first define the following entities that take part in the proposed architecture:

\begin{figure}[b]
\centering
\includegraphics[width=0.45\textwidth]{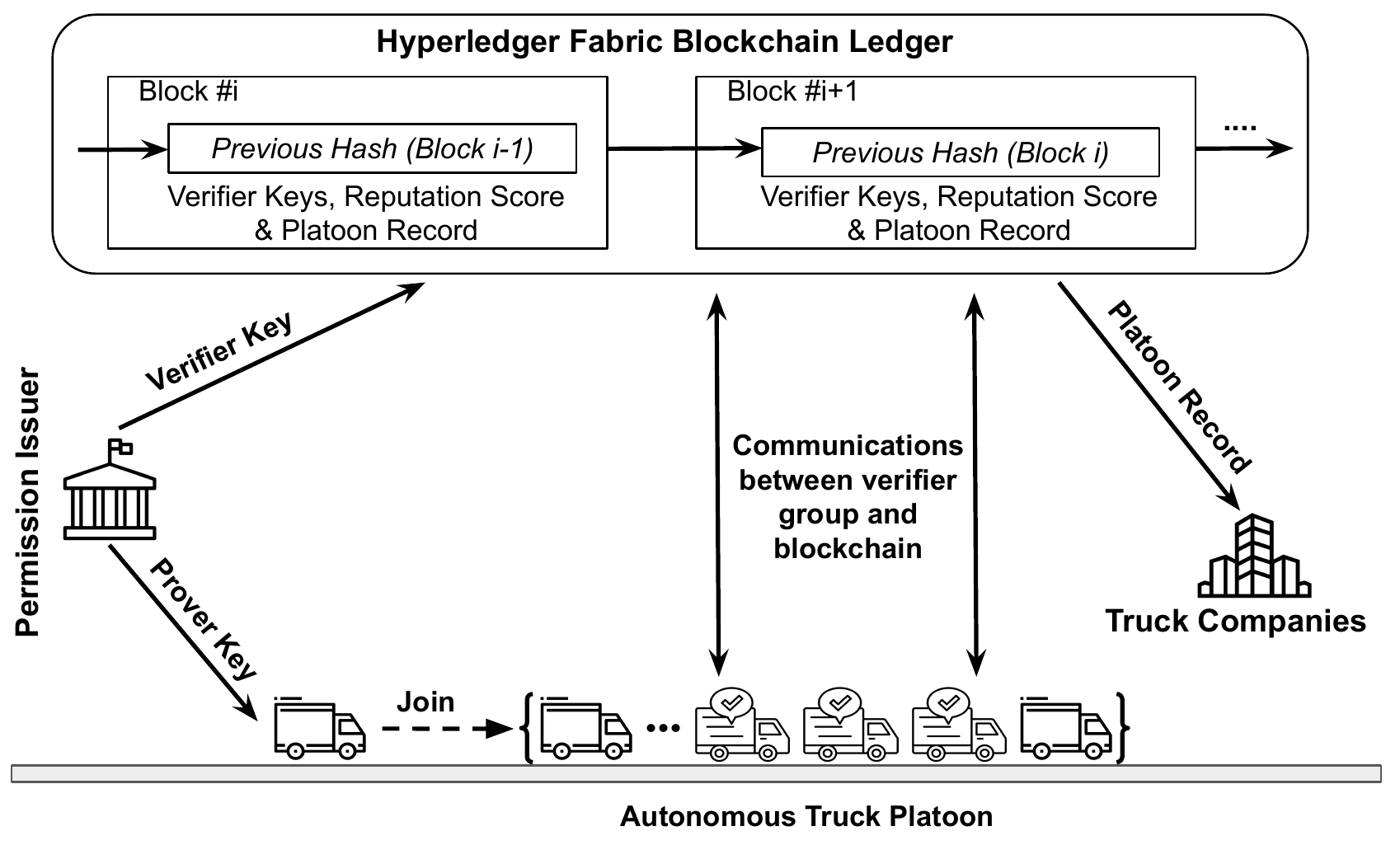}
\caption{Proposed System Architecture based on permissioned blockchain.}
\label{fig:arch}
\end{figure}

\begin{itemize}
    
    \item Permission Issuer: A permission issuer is a trusted entity who manages identifiers of autonomous vehicles (e.g., MAC address) and issue key pairs to data owners and data verifiers. In practice, an agency such as the Department of Motor Vehicles (DMV) can function as the permission issuer in the proposed system.
   
    \item Permissioned Blockchain: A permissioned blockchain (in our prototype using Hyperledger Fabric) is utilized as the controller of the architecture and serves as the tamper-proof transaction ledger for recording verifier keys, reputation scores and platoon records.
  
    \item Verifier Group: A verifier group is a subset of the platoon that includes the most trusted autonomous trucks. It validates the identities of new trucks before joining the platoon, and this group is dynamically updated based on the reputation scores of trucks in the current platoon. 
    
    \item Autonomous Truck: An autonomous truck is a participant of dynamic platoons and also a candidate for the spatially-local verifier group in the blockchain network. 
    
    \item Truck Company: A truck company is a client in the blockchain network who may need to retrieve its owned trucks' platooning histories. Trucking companies can use this information for practical applications, such as determining the optimal platoon size on each route to reduce fuel consumption or to quantify platooning benefits such as efficiency and safety improvements.
    
\end{itemize}

\subsection{Location-aware Verification Protocol}
%ZKP+Voting+Dynamic Verifier Group Formation
We introduce the location-aware verification protocol that validates the identity of the autonomous truck in a privacy-preserving manner. When an autonomous truck needs to join a platoon, it acts as the prover to prove its identity to the verifier group in ZKP-based proof. Trusted participants within the platoon form a spatially-local verifier group to validate the proof without revealing the information. The verifier group members are dynamically updated based on the sorting of their reputation scores.

\begin{theorem}
\label{theorem-1}
Let $G$ be a multiplicative cyclic group of prime order $p$ with generator $g$. Let $e : G \times G \rightarrow G_T$ be a computable, bilinear and non-degenerate pairing into the group $G_T$. Then, we have $e(x^a, y^b) = e(x, y)^{ab}$ for all $x,y \in G$ and $a,b \in \mathbb{Z}_p$ because $G$ is cyclic.
\end{theorem}

Based on Theorem \ref{theorem-1} \cite{cyclic}, we describe how to construct the location-aware verification protocol as shown in Algorithm \ref{algorithm3}. The event-driven algorithm mainly consists of four parts: lines 2-6 represent the phase of key generation; lines 7-16 start the joining request by the autonomous truck (prover) and show how zero-knowledge proof is generated; lines 17-31 verify the proof by the verifier group; and lines 32-41 update the verifier group after the autonomous truck joins the platoon. In addition to the following, Section \ref{zkp-experiment} details the prototype of our proposed ZKP scheme in the context of truck platooning. In the prototype, we choose BLS scheme \cite{boneh2001short} to build the generator $g$ and elliptic curve \cite{frey1999tate} for bilinear pairing $e$.

\begin{algorithm}[t]
\begin{algorithmic}[1]
\caption{Location-aware Verification Protocol}
\label{algorithm3}

\State \textbf{OUTPUT:} The identity verification result $r_i$ for prover $i$; 

\State \textbf{KEY GENERATION}
\State The permission issuer selects a random $a_i \in \mathbb{Z}_p$ and computes $v_i = g^{a_i} \in G$;
\State Prover key $a_i$;\Comment{Assign to prover $i$}
\State Verifier key $v_i = g^{a_i}$;\Comment{Save on blockchain}
\State \textbf{END KEY GENERATION} 

\State \textbf{START UP} {\em The prover $i$ (autonomous truck) starts a $request$ to join the platoon:}
\State The prover computes its hashed identity information $m_i$ in SHA256 algorithm \cite{rachmawati2018comparative}, as $h_i = H(m_i)$;
\State One-time zero-knowledge proof $\delta_i = {h_i}^{a_i} \in G$;
\State The prover sends $\delta_i$ to the verifier group;
\State The prover waits for $r_i$ in a time period $T$;
\State The prover gets $r_i$ in a time period $T$;
\If {no $r_i$ within time $T$} 
\State  Restart $request$;
\EndIf
\State \textbf{END START UP}

\State \textbf{UPON EVENT} {\em The verifier group receives the one-time zero-knowledge proof $\delta_i$:}
\For{each verifier $j$}
\If{$e(\delta_i, g) = e(h_i, v_i)$} \Comment{Theorem 1}
\State $r_{i,j} =$ TRUE;
\Else
\State $r_{i,j} =$ FALSE;
\EndIf
\EndFor
\State The verifier group returns the $r_i$ based on voting;
\If{$r_i ==$ TRUE}
\State Approve $request$;
\Else 
\State Reject $request$;
\EndIf
\State \textbf{END UPON EVENT}

\State \textbf{UPON EVENT} {\em The prover $i$ joins the platoon:}
\For{each verifier $j$}
\If{$r_{i,j} == r_i$}
\State $rs_j ++$; \Comment{Increase reputation score}
\Else
\State $rs_j --$; \Comment{Decrease reputation score}
\EndIf
\EndFor
\State Update verifier members based on sorting of each participant's reputation score;
\State \textbf{END UPON EVENT}

\end{algorithmic}
\end{algorithm}

\subsection{Blockchain Network% with Access Control Policy
}
Our permissioned blockchain system is prototyped using the Hyperledger Fabric platform. In our design, the blockchain functions as a distributed ledger which stores verifier keys, reputation scores, and truck platoon records. Data is maintained on-chain to guarantee immutability and integrity. By storing the platoon history records on-chain, we provide practical benefits to the fleet companies operating on our platform. For example, a trucking company can retrieve and analyze the platoon records for their vehicles in order to determine the optimal platoon size on each of their routes based on historical data. Furthermore, platooning provides additional benefits of efficiency and safety, and the platoon records stored on the blockchain can be leveraged to help a company quantify the benefits. %Due to the cost of data storage on blockchain, our design forgoes storing trucks vehicle information directly on the blockchain. Anyways, this data is already contained within the internal systems of the trucking company. 

\section{Experiments and Evaluation}
%login window, CAV
\subsection{Experimental Setup}
We prototype the proposed identity verification system and conduct a series of experiments to evaluate its performance. The system consists of two primary portions that interact seamlessly: the verification module based on ZKP and the blockchain network. The ZKP scheme is programmed by using the Hyperledger Ursa library \cite{hyperledgerursa}. The blockchain network is developed on the Hyperledger Fabric v1.2 and tested using the Hyperledger Caliper benchmark tool \cite{hyperledgercaliper}. For testing, we instantiate 10 participants, including 8 autonomous trucks and 2 companies, in the blockchain network. The prototype and experiments are deployed and conducted on multiple Fabric peers in Docker containers locally on Ubuntu 18.04 operating system with 2.8 GHz Intel i5-8400 processor and 8GB DDR4 memory.

\subsection{Verification Module Based on ZKP Scheme}
\label{zkp-experiment}
As illustrated in Fig. \ref{fig:zkp-process}, the ZKP scheme performs the functionalities of initial setup, generation and verification of zero-knowledge proofs of autonomous trucks' identifiers. These functionalities are programmed by using Hyperledger Ursa, a cryptographic library for Hyperledger applications. Hyperledger Ursa is programmed using the Rust language and provides APIs for various cryptographic schemes. Our ZKP module operates in the following three phases:

\begin{figure}[b]
\centering
\includegraphics[width=0.45\textwidth]{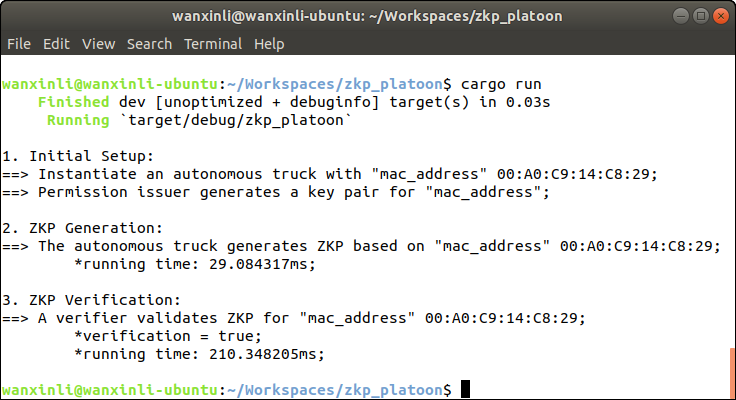}
\caption{Process of the ZKP scheme on Hyperledger Ursa.}
\label{fig:zkp-process}
\end{figure}

\subsubsection*{Phase 1 - Initial Setup}
Phase 1 initializes an autonomous truck instance to act as the prover. As shown in Fig. \ref{fig:zkp-process}, the truck has the identifier information {\tt mac\_address} (value: {\tt 00:A0:C9:14:C8:29}), and the permission issuer generates a key pair for the identity information. The BLS scheme \cite{boneh2001short} is used to build the key pair generator, which creates the prover key for the truck and the verifier key on the blockchain ledger, as follows:

\begin{verbatim}
let generator = Generator::new().unwrap();
let prover_key = SignKey::new().unwrap();
let verifier_key = VerKey::new(&generator,
                 &sign_key).unwrap();
\end{verbatim}

%The BLS scheme uses a bilinear pairing for verification, and signatures are elements of an elliptic curve group.

\subsubsection*{Phase 2 - ZKP Generation}
In this phase, the autonomous truck uses the prover key to generate a one-time zero-knowledge proof for the hashed {\tt mac\_address} via SHA256 algorithm \cite{rachmawati2018comparative}. The resulting proof consists of three elements on an elliptic curve. For instance, as shown below: 

\begin{figure}[h]
\centering
\includegraphics[width=0.49\textwidth]{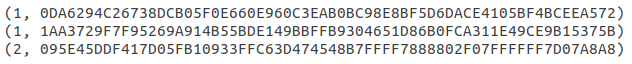}
%\caption{Example of generated zero-knowledge proof.}
\label{fig:proof-example}
\end{figure}

\noindent
the proof of hashed {\tt 00:A0:C9:14:C8:29} from {\tt mac\_address} is a combination of three points on an elliptic curve represented in hexadecimal format. Our experiments show that the average running time for the proof generation phase is 29 ms.  

\subsubsection*{Phase 3 - ZKP Verification}
After the proof generation, the verifier group from the platoon validate the zero-knowledge proof from the truck. The verification function takes the proof, the hashed {\tt mac\_address}, the verifier key and the corresponding generator as inputs, and utilizes elliptic curve bilinear pairing \cite{frey1999tate} to verify the proof:

\begin{verbatim}
let result = Bls::verify($proof, 
             mac_address.as_slice(),
             $verifier_key, $generator)
             .unwrap();
\end{verbatim}

In our experiments, the average running time for verifying each zero-knowledge proof is around 210 ms. After that, the platoon can authenticate the vehicle's identifier anonymously and, subsequently, communicate the result to both the truck and blockchain network.

\begin{figure}[b]
\centering
\includegraphics[width=0.45\textwidth]{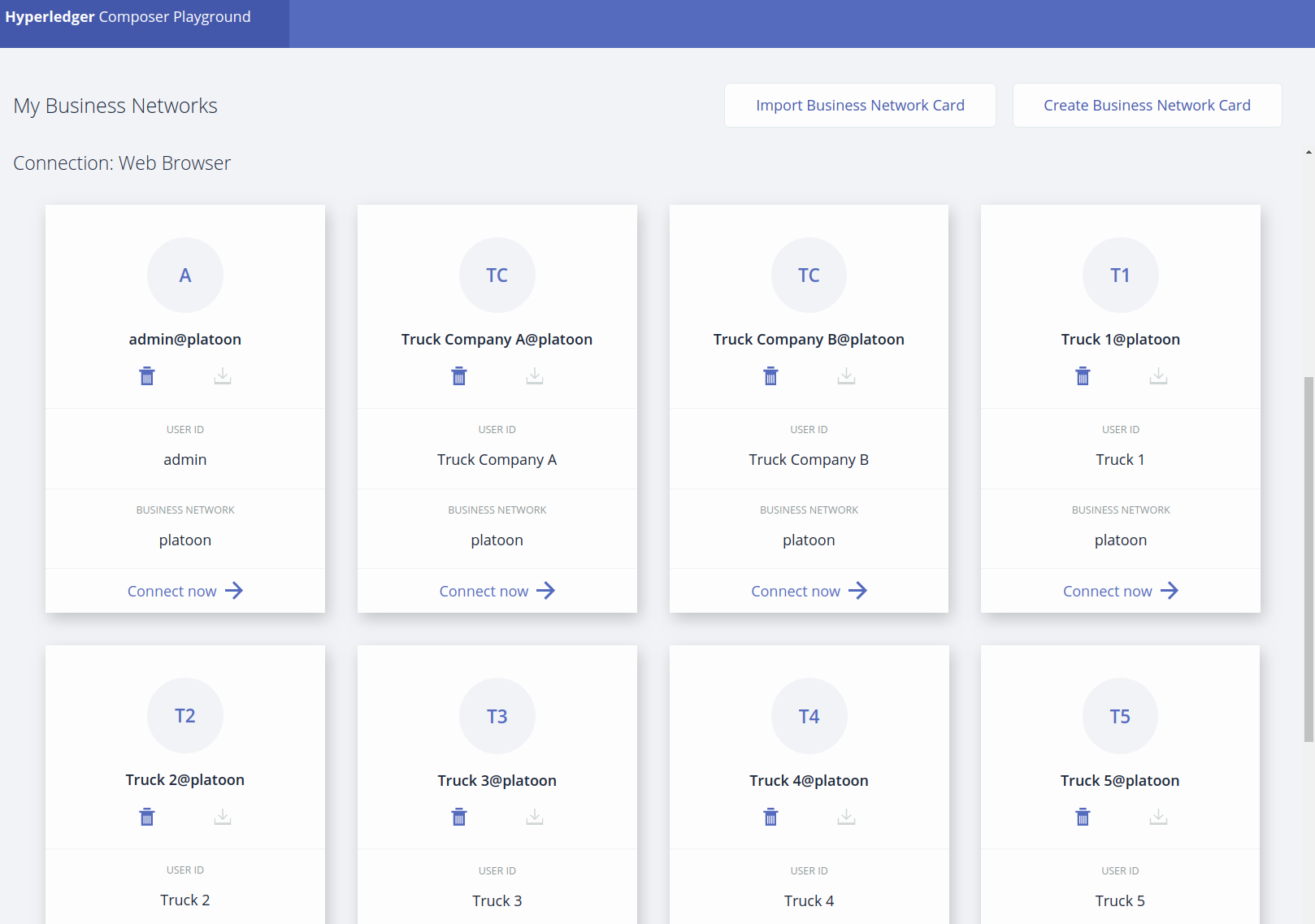}
\caption{Blockchain network login window for companies and autonomous trucks.}
\label{fig:login-window}
\end{figure}

\subsection{Blockchain Network}
\label{blockchain-network}
Hyperledger Fabric is an open-source and modular permissioned blockchain framework \cite{hyperledgerfabric}. The four programmable modules used in our system are: model file (.cto) which is used to define all of the data structures in the network; script file (.js) where smart contracts are written; access control list (.acl) for deploying access control policies; and the query file (.qry) which defines the query operations similarly to a traditional database system.

In our prototyped blockchain system, we provide a web portal for the network participants (autonomous trucks and companies), which can be used to interact with the blockchain network. An example can be seen in Fig. \ref{fig:login-window}, where each participant has a registered ID for connecting to the blockchain. The trusted entity operating as the permission issuer in our system (e.g., DMV) also acts as the blockchain administrator. %, issuing access control permissions for autonomous trucks and companies. 

\subsection{Performance Evaluation}

%\textcolor{blue}{Caliper Settings: Under each endorsement policy, there are 150 transactions in total during the test circle, and the send rate is set as 30 tps.}

\subsubsection{Transaction Throughput}
\begin{figure}[t]
\centering
\includegraphics[width=0.35\textwidth]{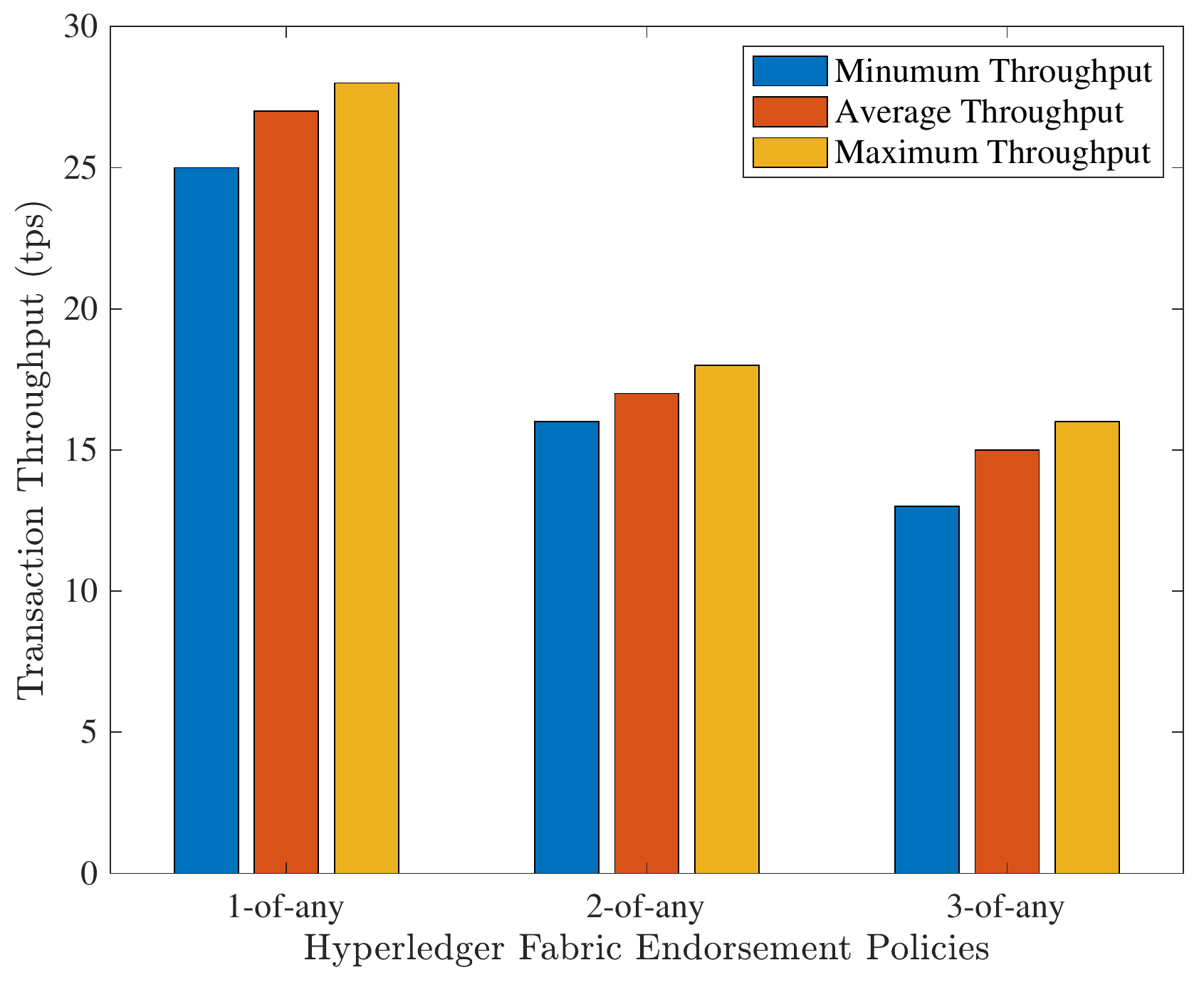}
\caption{Minimum, average and maximum transaction throughput vs. Hyperledger Fabric endorsement policy.}
\label{fig:tps_endorse}
\end{figure}

% \textcolor{blue}{Figure 5}
We first measure the transaction throughput of our blockchain network prototype. Transaction throughput for a blockchain network quantifies the rate at which transactions are processed through the network over a given time cycle in units of transactions per second. We tested the throughput under different endorsement policies, and the results for 1-of-any, 2-of-any, and 3-of-any policies are summarized in Fig. \ref{fig:tps_endorse}. Our results show that the number of endorsing peers has an inverse relationship to the transaction throughput of the network, and the transaction throughput peaks at 27 tps, 17 tps, and 15 tps under 1-of-any, 2-of-any, and 3-of-any policies respectively. 

The endorsement policy has a strong impact on transaction throughput. However, this makes sense because increasing the number of peers required to validate a transaction also increases the complexity of the endorsement process. That being said, our results from multiple rounds of testing show that for a given endorsement policy, the performance is relatively stable and the difference between the minimum, maximum and average cases is minor. 

\subsubsection{Transaction Latency}
\begin{figure}[t]
\centering
\includegraphics[width=0.35\textwidth]{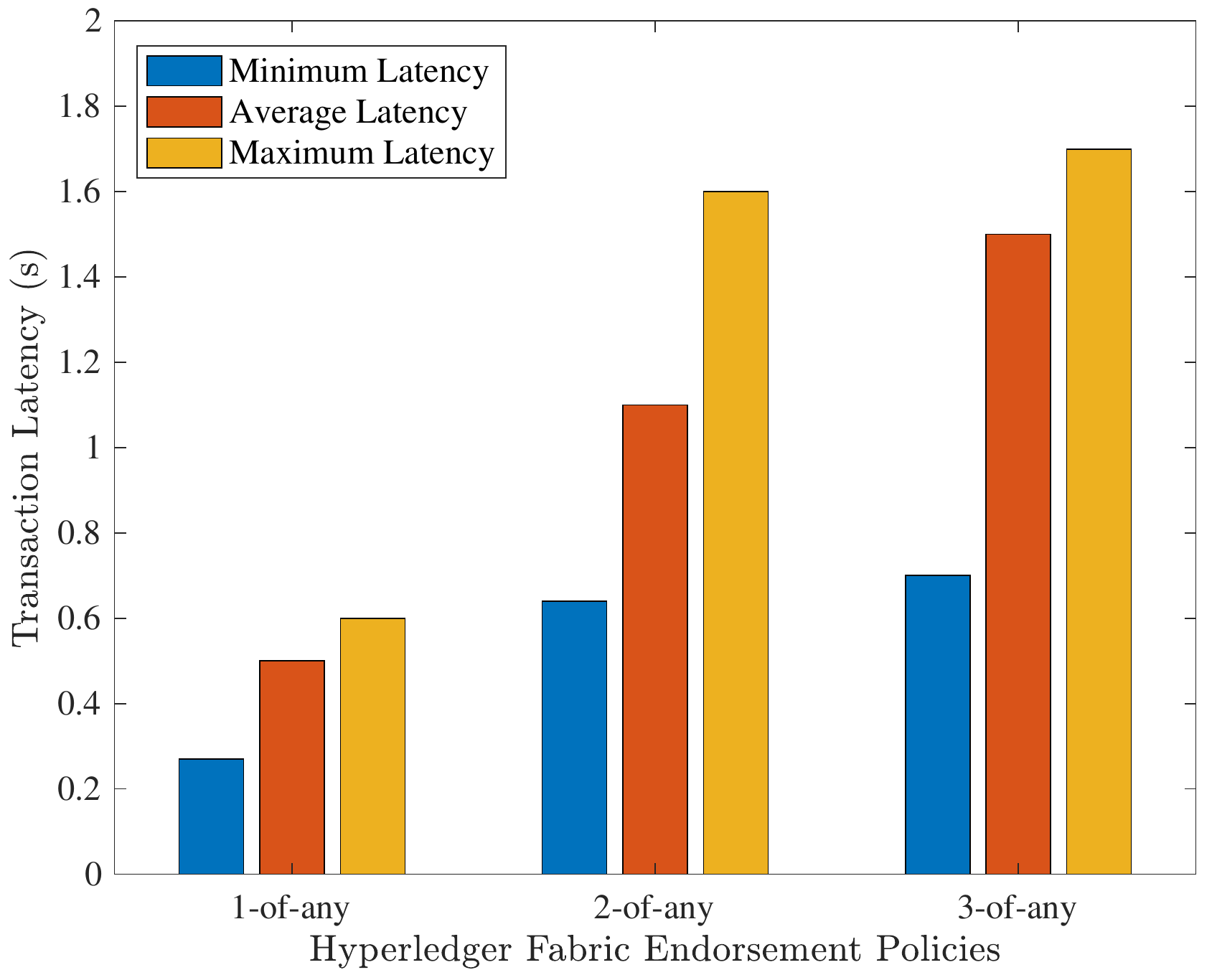}
\caption{Minimum, average and maximum transaction latency vs. Hyperledger Fabric endorsement policy.}
\label{fig:latency_endorse}
\end{figure}

% \textcolor{blue}{Figure 6}
We also perform experiments to measure and quantify the transaction latency of our prototyped blockchain network. Transaction latency measures the end-to-end processing time for a transaction in the blockchain network, from initial client submission to the time when the transaction is committed to the ledger. We perform multiple experiment rounds with varying endorsement policies and compiled our results in Fig. \ref{fig:latency_endorse}. The relationship between the transaction latency and endorsement policy is readily apparent: as the number of endorsing peers increases, we see an increase in both average and maximum transaction latencies. However, it is worth noting that as the number of endorsing peers increases, we also see an increase in variability between the minimum, average and maximum cases. That being said, the difference in latency between 2-of-any and 3-of-any endorsement policies is significantly lower than the difference between 1-of-any and 2-of-any cases. 

\section{Conclusion}
This paper introduces a novel location-aware and privacy-preserving verification protocol focused on the application of dynamic platooning for autonomous trucks. Our proposed system integrates zero-knowledge proof with permissioned blockchain technology. By performing the ZKP-based identity verification within the spatially-local area defined by a given platoon, our system can provide lower latency verification with less communication overhead compared to a location-agnostic system. %Platoon records and reputation scores are stored directly on-chain to guarantee immutability and integrity of the data. Privacy for stored data is ensured through our programmable access control policies, which enable a trucking company to define who can access their platoon records. 
To analyze the system performance, we prototype our design on Hyperledger platform and perform various experiments. Initial results highlight our system's real-world feasibility for providing both low-latency identity verification and transaction processing on the order of milliseconds. %Additionally, we discuss the security of our system against possible attack vectors, such as the platoon record tampering and fake autonomous truck attacks, to highlight its robustness. For future research, we plan to study a detailed incentive mechanism for calculating truck reputation scores, as well as investigating location-focused group formation mechanisms. %directly within the consensus layer. 

%In future works, we plan to integrate the verification process directly into the consensus layer, as well as experiment with permissionless blockchain systems.

% % use section* for acknowledgment
% \section*{Acknowledgment}




%\bibliographystyle{IEEEtran}


% \end{thebibliography}

\Urlmuskip=0mu plus 1mu\relax

\bibliographystyle{IEEEtran}
\bibliography{sigproc.bib}

\end{document}